\documentclass[pre,twocolumn,superscriptaddress,floatfix,showpacs]{revtex4}

\usepackage[latin2]{inputenc}
\usepackage{amsmath}
\usepackage{amssymb}
\usepackage{epsfig}
\usepackage{color}
\usepackage{bm}
\usepackage{cancel}
\begin{document}

\title{Stripe formation in horizontally oscillating granular suspensions}

\author{Robabeh Moosavi}
\affiliation{Department of Physics, Institute for Advanced 
Studies in Basic Sciences, Zanjan 45137-66731, Iran}

\author{Maniya Maleki}
\email{m_maleki@iasbs.ac.ir}
\affiliation{Department of Physics, Institute for Advanced 
Studies in Basic Sciences, Zanjan 45137-66731, Iran}

\author{M.\ Reza Shaebani}
\email{shaebani@lusi.uni-sb.de}
\affiliation{Department of Theoretical Physics, Saarland 
University, D-66041 Saarbr\"ucken, Germany}

\author{J. C. Ruiz-Su\'arez}
\affiliation{CINVESTAV-Monterrey, PIIT, Apodaca, Nuevo Leon, 
66600, Mexico}

\author{Eric Cl\'ement}
\affiliation{PMMH, ESPCI, UMR CNRS 7636 and Universit\'e Paris 
6 et Paris 7, 75005 Paris, France}

\date{\today}

\begin{abstract}
We present the results of an experimental study 
of pattern formation in horizontally oscillating granular 
suspensions. Starting from a homogeneous state, the suspension 
turns into a striped pattern within a specific range of 
frequencies and amplitudes of oscillation. We observe an 
initial development of layered structures perpendicular 
to the vibration direction and a gradual coarsening of the 
stripes. However, both processes gradually slow down and 
eventually saturate. The probability distribution of the 
stripe width $P(w)$ approaches a nonmonotonic steady-state 
form which can be approximated by a Poisson distribution. 
We observe similar structures in MD simulations of soft 
spherical particles coupled to the motion of the surrounding 
fluid.
\end{abstract}


\maketitle

\section{Introduction}
Granular mixtures and suspensions exhibit a rich variety 
of behavior, including segregation, clustering, and pattern 
formation, when subjected to mechanical agitation 
\cite{Aranson06,Kudrolli04}. Because of the ubiquity of 
such phenomena in nature and industry, understanding the 
mechanisms behind the instability of homogeneous state 
in granular systems is of great interest within the physics 
and engineering communities. When a mixture of dry grains, 
differing in shape, size, density, or other physical 
properties is agitated, separation between different 
species may occur depending on the properties of the 
external drive. In the case of strong driving the 
dynamics is dominated by inelastic binary collisions. 
Here, effective long-range interactions appear between 
the particles of the same type due to the presence of 
hydrodynamic fluctuations \cite{Sanders04,LongRangeForces}, 
leading to a variety of collective behaviors. With 
decreasing the driving strength, durable contacts form 
between the particles, eventually becoming dominant. 
In this regime, friction plays a crucial role 
\cite{Tennakoon98,Krengel13} and relevant separation 
mechanisms are e.g.\ local rearrangements, percolation, 
and convection \cite{DurableContMech}. The dynamics is 
more complex in granular suspensions due to the influence 
of the fluid on the particles, causing e.g.\ anisotropic 
hydrodynamic forces or viscous drag. The existence of 
underlying hydrodynamic interactions is the reason 
behind different time scales for pattern formation 
in suspensions compared with dry mixtures.
 
\begin{figure}[b]
\centering
\includegraphics[trim=3cm 0cm 0cm 0.5cm, clip=true, 
width=0.9\columnwidth]{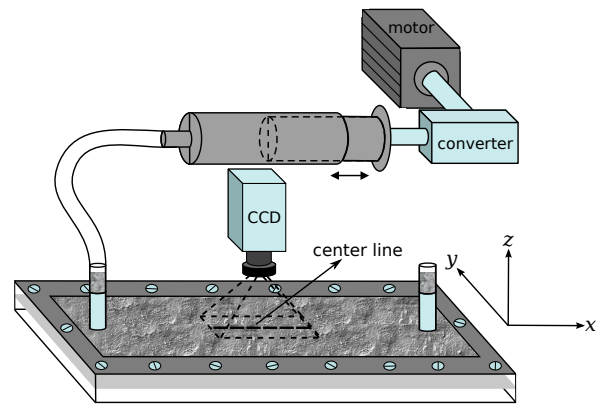}
\caption{Schematic picture of the experimental setup.}
\label{Fig1}
\end{figure}

The observed behavior also depends on the agitating method 
by which energy is injected into the system. For example, 
vertical vibrations have attracted a lot of attention since 
they can lead to variants of the Brazil nut effect 
\cite{Clement10,BrazilNut} or clustering \cite{Sanders04,
VerticalClustering}. The dynamics of horizontally driven 
systems is even more diverse. Shear-induced segregation 
has been reported in suspensions \cite{Barentin04} and 
granular mixtures in a split-bottom cell \cite{Hill08}, 
where the presence of shear bands \cite{ShearBand} 
renders the system inhomogeneous. Horizontal swirling 
motion may also lead to separation of
\begin{figure*}
\center
\includegraphics[width=0.9\textwidth]{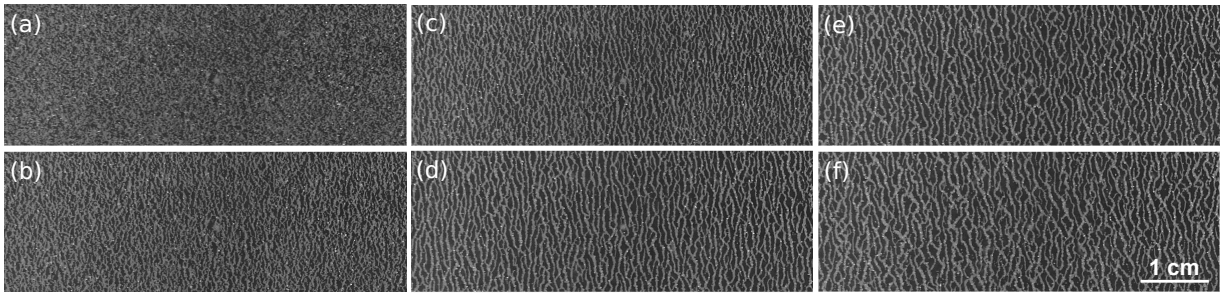}
\caption{Evolution of the stripe patterns in a suspension 
of polystyrene beads (bright) immersed in a NaCl aqueous 
solution (dark), which is vibrated at $f {=} 19.8 \, 
\text{Hz}$ and amplitude of vibration $A {=} 0.21 \, 
\text{mm}$. Snapshots are taken at $t {=} 4$, $12$, $20$, 
$52$, $104$ and $340 \, \text{seconds}$, from (a) to (f), 
respectively.}
\label{Fig2}
\end{figure*} 
different species \cite{Swirling}. 
One of the most interesting types of instability is the 
formation of stripes, which has been observed in horizontal 
shaking or oscillatory excitation of binary mixtures of dry 
particles \cite{Reis02,Reis04,Reis06,PicaCiamarra05,Mullin00,
Pooley04,Fujii12,Ehrhardt05,PicaCiamarra06}, or a single 
type of dry particles with a moderate size polydispersity 
\cite{Krengel13}. There have been less studies on pattern 
formation in suspensions. S\'anchez et al.\ \cite{Sanchez04} 
investigated the emerging striped patterns in a 
fluid-immersed mixture of bronze and glass spheres subjected 
to horizontal vibration, and proposed a mechanism for the 
formation of stripes based on the differential influence 
of drag on the components of the mixture. However, the 
development of chain-like structures has been observed 
even between identical spheres in oscillatory fluid flows 
in MD simulations \cite{Klotsa09}. 

In this Letter, we study the formation of stripes in a 
suspension consisting of a single type of particles 
immersed in water, which is subjected to horizontal 
sinusoidal vibration. We elucidate how the striped 
pattern evolves in time by considering the structure 
at the particle level. The focus of research so far 
has been on identifying the criteria under which the 
stripes emerge. It is known that no pattern forms at 
extremely low or high frequencies and amplitudes of 
vibration \cite{Sanchez04,Reis06,Krengel13} or below 
a certain level of the density of particles \cite{Reis02,
Ehrhardt05,Reis06}. On the other hand, within the 
appropriate range of parameters, stripe formation is 
a remarkably dominant process. For example, even initially 
separated phases exhibit shear-induced instabilities 
under oscillatory excitation \cite{PicaCiamarra05,
PicaCiamarra06,Maleki07} and ultimately end up with 
stripe-like structures. However, the structural properties 
of the patterns and their spatial and temporal evolution 
is poorly investigated and understood. It has been 
shown that a gradual coarsening of the structure occurs 
\cite{Mullin00,Reis02,Sanchez04,Ehrhardt05,Reis06,Mullin13}, 
and that the average width of the stripes increases 
with increasing the particle density \cite{Reis02,
Ehrhardt05,Reis06}. Here, we characterize the structure 
by measuring the thickness $w$ of the stripes. By 
monitoring the time evolution of the stripe-width 
probability distribution $P(w)$ at a given particle 
density, we verify that the coarsening procedure 
finally stops and $P(w)$ tends towards a steady 
state nonmonotonic distribution with a peak 
around $3$ particle-diameter thickness. Besides 
coarsening, our results evidence the existence 
of another underlying dynamical process, which 
simultaneously arranges the particles in vertical 
layers oriented in the vibration direction. MD 
simulations of intruder particles immersed 
in a viscous fluid reproduces the experimentally 
observed patterns.

\section{Experimental setup}
Our experimental setup is a Hele-Shaw cell consisting 
of two parallel plexiglass plates of lateral sizes $20
\text{cm} {\times} 5\text{cm}$ separated by a distance 
$s {=} 400\,\mu \text{m}$. The gap $s$ is accurately 
controlled by using a Teflon sealed Mylar sheet which 
is sandwiched between the plates along the perimeter 
of the cell by small screws. A schematic picture of the 
setup is shown in Fig.~\ref{Fig1}. The cell is filled 
with a suspension which contains either polystyrene or 
glass beads immersed in water. Although similar patterns 
form in both cases, here we only present the results 
of the polystyrene suspension. The beads are spherical 
with diameter $d {=} 80\,\mu \text{m}$ and density $\rho 
{=}1.05 \, \text{g}/\text{cm}^3$ and occupy a volume 
fraction of $7\%$. To avoid electrostatic effects, we 
use a NaCl aqueous solution with a density of $\rho_s 
{=} 1.101 \, \text{g}/\text{cm}^3$.

Two inlet and outlet holes are created in the top 
plate close to the two ends of the cell along the 
$x$ direction (see Fig.~\ref{Fig1}). The outlet hole 
is exposed to the open air, and the inlet hole is 
connected to a syringe by means of a soft silicone 
tube. The syringe is attached to a mechanical 
converter which transforms the rotational motion of 
an AC motor into a sinusoidal vibration. This periodic 
motion is transferred to the syringe and induces a 
back and forth motion of the suspension inside the 
tube and cell. This is a novel method to produce 
oscillatory motion by vibrating the suspension 
itself instead of the container. The frequency of 
the vibrations is controlled by an AC speed controller, 
and the amplitude is held constant at $A {=} 0.21 \, 
\text{mm}$. The cell is filled with suspension, 
without allowing visible air bubbles to be trapped 
inside. 

We image the suspension from above with a CCD camera 
with a resolution of $40\,\mu\text{m}/\text{pixel}$.
The resulting images are indeed the projections of 
particle configurations into the $x{-}y$ plane. 
The camera zooms in on a region of size $6\text{cm}
{\times}2\text{cm}$ around the center of the cell 
(see Fig.~\ref{Fig1}), where the flow field remains 
uniform far from the boundaries. In order to 
characterize the structural properties of the patterns, 
the images can be rather easily analyzed since the 
contrast between the solution and the grains gives 
an opportunity to clearly distinguish between them. 
Thus a simple thresholding converts the gray-scale 
images into binary ones, with $1$ (white) and $0$ 
(black) pixels denoting the grains and the solution, 
respectively.

\begin{figure}
\center
\includegraphics[width=0.45\textwidth]{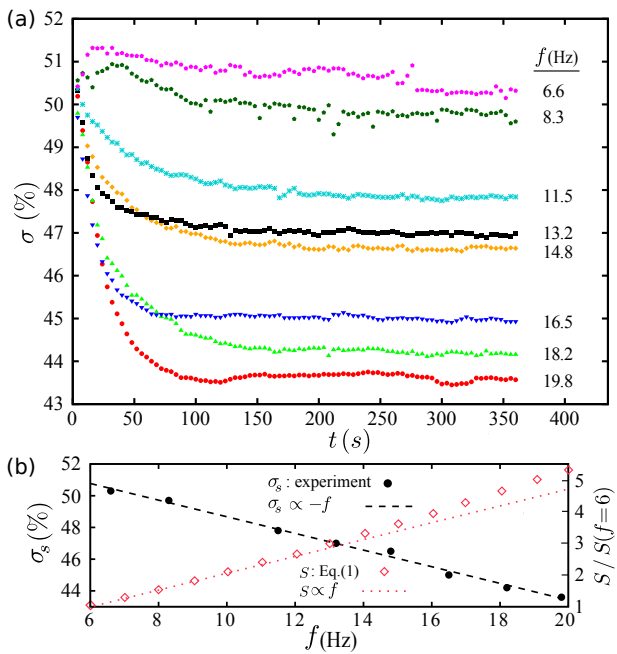}
\caption{(a) The area fraction $\sigma$ occupied by 
grains as a function of time for a suspension vibrated 
with amplitude $A {=} 0.21\,\text{mm}$ at different 
frequencies $f$. (b) The stationary value of the 
occupied area fraction $\sigma_{\!s}$ and the average 
Shields parameter $S$ versus the vibration frequency $f$.}
\label{Fig3}
\end{figure}

\section{Formation of stripes}
Starting from a well-mixed initial condition (with 
density fluctuations below $5\%$ on grids of one hundred 
particle-diameter size), the homogeneous state becomes 
unstable within the first few seconds, and stripe patterns 
emerge which span the entire area of the cell. Clear 
stripe formation is observed within the moderate frequency 
range of $6{-}20 \, \text{Hz}$, while no pattern forms below 
a volume fraction of $4\%$ or above $15\%$ . However, we do not aim 
here at exploring the frequency-amplitude space for the 
pattern formation subdomain. Figure \ref{Fig2} shows 
the time development of the system under horizontal 
sinusoidal vibration at frequency $f{=}19.8\,\text{Hz}$. 
The stripes form nearly perpendicular to the 
oscillation direction and exhibit a branched structure 
with a persistence length of a few millimeters (i.e.\ 
a few tens of particle diameters). A closer look at 
the evolution of the system interestingly indicates 
the presence of two underlying dynamical processes: (i) 
development of layered structures, and (ii) coarsening. 
The freely floating particles and/or the randomly 
formed stripes organize themselves into layers along 
the vertical axis perpendicular to the vibration 
direction on one hand, and merge into thicker horizontal 
bands on the other hand. 

Let us first investigate the ordering of particles 
in layers. If the particles sit on top of each other 
along the $z$ direction, they should occupy a smaller area 
fraction of the cell when observed from the top view. 
Note that the thickness of the cell allows the formation 
of layers with heights of up to $5$ particle diameters. 
We measure the occupied area fraction $\sigma$ (i.e.\ 
the ratio between the number of white pixels and the 
total number of pixels in a picture) during the experiment 
for several values of the oscillation frequency. While 
$\sigma$ remains unchanged or quickly (in a few seconds) 
reaches a steady value at low frequencies, it initially 
decreases with time for high frequencies. The results 
shown in Fig.~\ref{Fig3}(a) however reveal that the slope 
gradually decreases, and $\sigma$ eventually saturates 
towards a stationary value $\sigma_{\!s}$ over a 
characteristic time $\tau_{_{\!l}}$. We observe no 
clear dependence of $\tau_{_{\!l}}$ on $f$, but the 
saturation value $\sigma_{\!s}$ approximately linearly 
decreases with increasing $f$ [see Fig.~\ref{Fig3}(b)], 
indicating that the ordering of particles in layers 
is more pronounced at higher frequencies. A plausible 
scenario is that a higher $f$ induces a stronger shear 
stress $\tau$ (${\sim}\partial v/\partial z$) which 
increases the resuspension of sedimenting particles 
(reflected in the increase of the Shields parameter 
$S{\sim}\tau$) and, thus, enhances the layering effect. 
To estimate $S(f)$, we solve Navier-Stokes and 
continuity equations for the fluid flow between two 
infinite parallel plates (i.e.\ far from the boundaries 
in the $x{-}y$ plane), leading to the diffusion equation 
$\frac{\partial v_x}{\partial t}{=} \nu \frac{\partial^2 
v_x}{\partial z^2}$, with $\nu$ being the kinematic 
viscosity. The solution should fulfill the boundary 
conditions $v_x(z{=}0,t) {=} v_x(z{=}L_z,t) {=} 0$ 
and $v_x(z{=}\frac{L_z}{2},t) {=} A \omega \, 
\text{cos}(\omega t)$. For simplicity, we take the 
solution in the limit of $\sqrt{2\nu/\omega}{\ll}L_z$ 
and obtain the average Shields parameter over 
one period of oscillation and an arbitrary sediment 
layer $\Delta$ as
\begin{equation}
S(f) {\sim} \frac{1}{\Delta}\!\int_0^\Delta \!\!\!\!f 
\!\!\int_0^{\frac{1}{f}} 
\!\! \Big|\frac{\partial v(z,t)}{\partial z}\Big| \, 
\text{d}t \, \text{d}z {=} \frac{2\pi f}{\Delta} 
\big(1{-}e^{-\Delta\sqrt{\frac{\pi f}{\nu}}}\big).
\label{Eq:ShieldsParameter}
\end{equation}
As shown in Fig.~\ref{Fig3}(b), $S$ grows slightly 
faster than linear with $f$ (with an exponent 
$1{<}\alpha{<}3{/}2$), which leads to a more 
pronounced ordering of particles in layers. 

\begin{figure}[t]
\centering
\includegraphics[width=0.9\columnwidth]{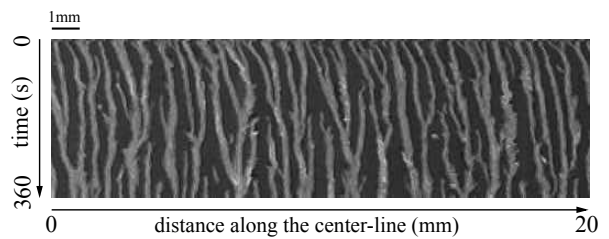}
\caption{Time-lapse image of the patterns formed 
along the center-line of the cell, at the same 
parameter values as in Fig.~\ref{Fig2}.}
\label{Fig4}
\end{figure}

Next, we address the coarsening behavior. 
Figure~\ref{Fig2} shows that the stripes gradually 
coarsen, i.e.\ thin stripes merge and form thicker 
ones. However, the stripes may also shrink, dissolve, 
or even branch to smaller stripes. The competition 
eventually leads to a dynamical saturation where 
the overall shape of the pattern becomes time 
invariant. Coarsening has been reported for dry 
\cite{Mullin00,Reis02,Ehrhardt05,Reis06} or 
fluid-immersed \cite{Sanchez04} binary mixtures 
of grains. The process is more clearly visualized 
in Fig.~\ref{Fig4}, where the time evolution of a 
specific part of the cell is displayed. To this aim, 
we focused on an imaginary line segment around the 
middle of the cell oriented along the vibration 
direction (``center-line" in Fig.~\ref{Fig1}). 
This line intersects many stripes which evolve in 
time. By recording a time-lapse sequence of frames 
over a few minutes, it is shown in Fig.~\ref{Fig4} 
that the mean number of stripes along the center-line 
initially decreases while their thicknesses grow on 
average. However, a steady state is reached after a 
characteristic time $\tau_{_{\!c}}$. 

\begin{figure}[t]
\centering
\includegraphics[width=0.9\columnwidth]{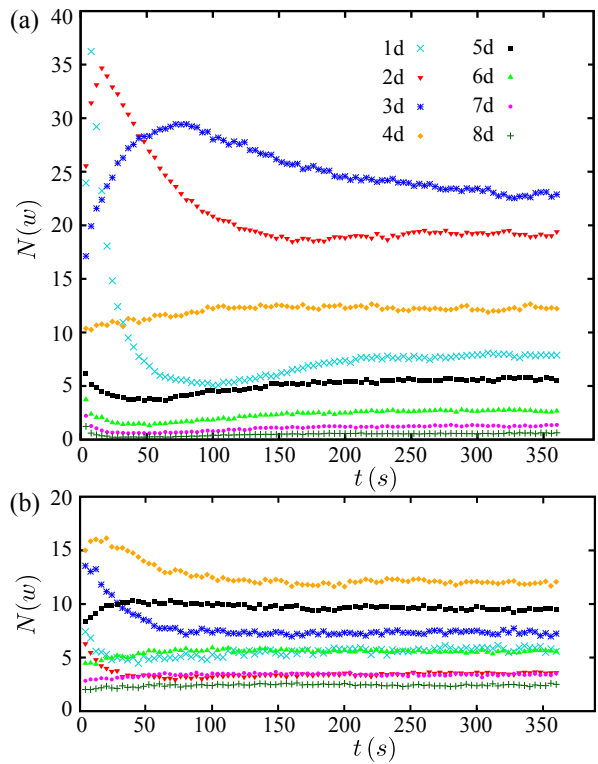}
\caption{The average width-resolved number of stripes 
$N(w)$ along the $x$-axis versus time, at frequency 
(a) $f {=} 19.8 \, \text{Hz}$ and (b) $f {=} 6.6 \, 
\text{Hz}$. The results are separately shown for 
several values of $w$, denoted in the units of grain 
diameter $d$.}
\label{Fig5}
\end{figure}

We would like to emphasize that the nature of layering 
and coarsening processes are different, thus, they 
develop on different time scales. For example, one 
obtains $\tau_{_{\!l}} {\sim}100 \,\text{s}$ vs.\ 
$\tau_{_{\!c}}{\sim}200 \,\text{s}$ at frequency $f 
{=} 19.8 \, \text{Hz}$. Moreover, the layering 
mechanism is fully suppressed in a quasi two-dimensional 
system, while the coarsening continues to occur. 
Another point is that both processes reach a dynamical 
saturation, meaning that the stripes continuously form, 
merge, or dissolve in time, even at $t {>} \tau_{_{\!l}}, 
\tau_{_{\!c}} $, only the relevant statistical 
measures, like the occupied area fraction $\sigma$ 
or the average stripe thickness, remain almost 
unchanged.

In order to quantitatively describe the coarsening 
procedure we categorize the stripes in terms of their 
width $w$ (by rounding off $w/d$ to integer numbers), 
and determine the number of stripes $N(w)$ in each 
category. This is done for each frame by scanning 
through each of the few hundreds of horizontal lines 
along the oscillation direction. In Fig.~\ref{Fig5}(a) 
we show the time evolution of $N(w)$ for several 
values of thickness $w$ at high frequency $f {=} 19.8 
\,\text{Hz}$. Interestingly, it turns out that $N$ 
evolves differently in time for different values 
of $w$. Apart from an initial transient, $N(w)$ 
decreases rather fast for thin stripes ($w/d {\simeq} 
1,\, 2$) while it considerably increases with time 
for intermediate values of thickness ($w/d {\simeq} 
3$) with a peak value around $t{\sim}70 \,\text{s}$. 
Comparison with thick stripes reveals only very 
modest changes in $N(w)$ for $w/d{\geq}4$. Moreover, 
the curves asymptotically converge to different 
values with a noticeable nonmonotonic ordering so 
that the stripes of width $w_{\!_\text{max}}{\sim}
3d$ are the most frequent ones at long times.

\begin{figure}[t]
\center
\includegraphics[width=0.9\columnwidth]{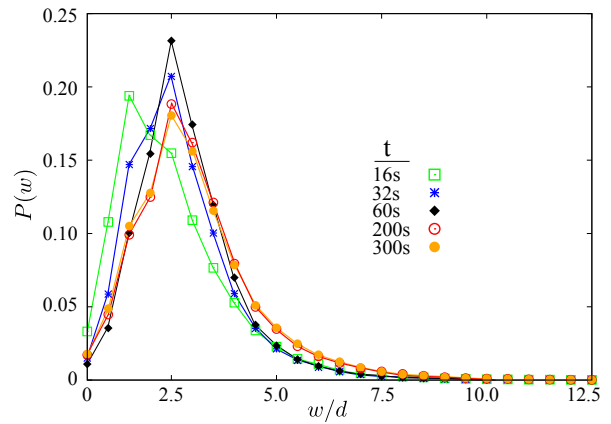}
\caption{Evolution of the probability distribution 
of the stripe width $P(w)$, obtained from 
experiments at $f {=} 19.8 \,\text{Hz}$.}
\label{Fig6}
\end{figure}

A comparison between the results at high and low 
values of $f$ reveals that the behavior also depends 
on the vibration frequency. In Fig.~\ref{Fig5}(b) we 
present the results for the low frequency $f {=} 6.6 \, 
\text{Hz}$. One observes that there are less stripes 
formed in the system compared with Fig.~\ref{Fig5}(a), 
and that the variation range of $N(w)$ is relatively 
small for all values of $w$, i.e.\ the initially formed 
patterns do not change much with time. Therefore, 
lowering of frequency slows down the underlying dynamics 
of the evolving patterns, reflected in the total number 
of stripes, in the coarsening behavior, and in the 
formation of layers (Fig.~\ref{Fig3}). At extremely 
low frequencies no pattern forms and the particles 
move with the surrounding fluid in a periodic manner. 
Figure \ref{Fig5}(b) also shows that the dominant 
steady-state width shifts towards thicker stripes 
($w_{\!_\text{max}}{\sim}4d$) with decreasing $f$. 
Roughly speaking, the relaxation time of an object 
of size $w$ in the fluid flow (${\sim}w^2$ in low 
Reynolds number regime) tends to be synchronized with 
the characteristic time of the fluid motion (${\sim}
1{/}f$) in the steady state, which results in 
$w{\sim}1{/}\sqrt{f}$. However, in practice, the 
most probable width $w_{\!_\text{max}}$ increases 
slower with decreasing $f$.

\section{Steady state} 
Next we investigate the probability distribution 
$P(w)$ of the stripe thickness to better understand 
the evolution of patterns. While the initial state 
of the patterns depends on the preparation 
conditions, after a short transient time $t_s$ 
(of the order a few seconds), $P(w)$ starts to 
systematically evolve towards the steady state 
distribution $P_{\!_\infty}\!(w)$: First, $P(w)$ 
becomes narrower while the peak position shifts 
towards wider stripes. Nevertheless, this process 
stops after a while, then, the peak position 
remains unchanged whereas $P(w)$ gets gradually 
broader and relaxes towards $P_{\!_\infty}\!(w)$ 
(see Fig.~\ref{Fig6}).

\begin{figure}[t]
\center
\includegraphics[width=0.9\columnwidth]{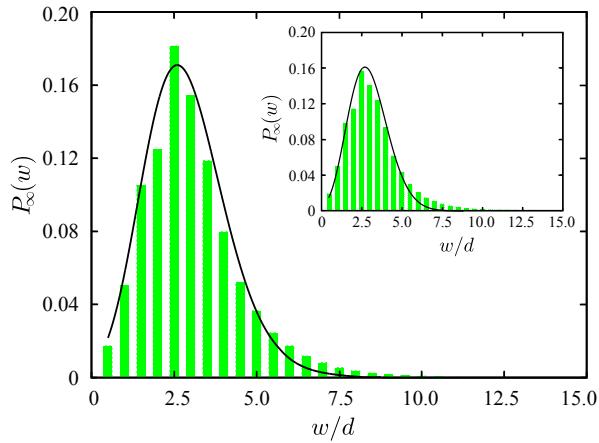}
\caption{$P_{\!_\infty}\!(w)$ at $f {=} 19.8 \,\text{Hz}$ 
in the steady state ($t{=}320s$). The solid line denotes 
a fit by Eq.~\ref{Eq:ProbDistInf} with ${\bar w}{=}2.8d$. 
Inset: $P_{\!_\infty}\!(w)$ at $f {=} 16.5 \,\text{Hz}$.}
\label{Fig7}
\end{figure}

Let us denote the mean number of particles on a 
horizontal plane and the mean number of stripes 
along the  oscillation direction (i.e.\ the 
$x$-axis) by $\bar n_\text{xy}$ and $\bar N$, 
respectively. Since both $\bar n_\text{xy}$ 
[according to Fig.~\ref{Fig3}(a)] and $\bar N$ (not 
shown) reach a saturation value in the steady state, 
the mean number of particles on a 1D band parallel 
to the $x$-axis, $\bar n_\text{x}{=}\bar n_\text{xy}
{/}(L_\text{y}{/}d)$, and thus the average stripe-width 
$\bar w{=}\bar n_\text{x}{/}\bar N$ converge to 
constant values as well. Assuming that, far from the 
initial conditions, the evolution of a stripe (including 
formation, coarsening, shrinkage, or dissolving) is 
a purely stochastic process, we consider the stripe 
width $w$ as a random variable in the interval $[0,L_x]$ 
with the known average $\bar w$. One expects that 
under these conditions, the steady-state probability 
distribution of $w$ can be approximated by the 
Poisson distribution: 
\begin{equation}
P_{\!_\infty}\!(w) = \displaystyle\frac{{\bar w}^w 
\text{exp}(-\bar w)}{ w!}.
\label{Eq:ProbDistInf}
\end{equation}
A quantitative comparison between the above equation 
and the experimental data is presented in Fig.~\ref{Fig7}. 
The overall agreement is satisfactory, even though 
the tail of the experimental data decays slower than 
the Poisson function. We note that deviation of 
$P_{\!_\infty}\!(w)$ from Eq.~(\ref{Eq:ProbDistInf}) 
occurs at low frequencies of oscillation. In the case 
of binary dry mixtures, a Gaussian distribution was 
reported for $P_{\!_\infty}\!(w)$ whose mean depends 
on the occupied volume fraction \cite{Reis02}. 

\section{Simulation results} 
To aid our understanding of the mechanism of the stripe 
formation, we perform molecular dynamics (MD) simulations, 
which has been previously employed to reproduce the 
experimental data \cite{Sanchez04,PicaCiamarra05,Krengel13}. 
We consider spherical soft particles with radii uniformly 
distributed between $0.95\bar r$ and $1.05\bar r$, which 
occupy a volume fraction of $\phi{\simeq}0.11$. The 
simulation box has dimensions $L_x {\times} L_y {\times} 
L_z$ (with $L_x {=} 10^4\bar r$, $L_y {=}10^3\bar r$, 
and $L_z {=} 10\bar r$) and is periodic in the $x$ 
direction. The ratio between the particle and the fluid 
density is set to $0.96$. The normal component of the 
interparticle interaction is given by a Hertzian 
repulsive force with viscous dissipation, 
\begin{equation}
F_n = k\delta^{3/2}+\gamma_{_n} \delta^{1/2}\dot \delta,
\label{Eq:Fn}
\end{equation}
where $k$, $\delta$, and $\gamma_{_n}$ are the spring constant, 
normal overlap, and damping parameter, respectively. The 
particle-wall interaction is treated in the same way. $k$ 
and $\gamma_{_n}$ depend on the elastic properties of the 
material and the radii of colliding particles (in practice, 
the average radius at low size polydispersity) \cite{Brilliantov96}. 
The interaction in the tangential direction is modeled 
using the force law $F_t {=} \gamma_{_t} |v_t^\text{rel}|$, 
with $\gamma_{_t}$ and $v_t^\text{rel}$ being the viscous 
damping and the relative surface velocity in the tangential 
direction, respectively. The Coulomb's criterion constrains 
the upper limit of $F_t$ to $F_t^\text{max} {=} \mu |F_n|$ 
(The data is shown for the friction coefficient $\mu{=}0.5$). 
The fluid-particle interaction is modeled via an effective 
Stokes's drag force $F_d$ proportional to the relative 
velocity $v_f^\text{rel}$ between the particle and the 
fluid, $F_d{=}{-}\gamma_{_d} v_f^\text{rel}$ \cite{Sanchez04,
PicaCiamarra05,PicaCiamarra06}, where $\gamma_{_d}$ (obtained 
from the empirical equation of Ergun \cite{Ergun52}) depends 
on viscosity and volume fraction. The influence of thermal 
fluctuations can be incorporated by adding a random noise 
to this equation (resulting in a Langevin-type dynamics) 
\cite{Lowen}.  

\begin{figure}[t]
\center
\includegraphics[width=0.9\columnwidth]{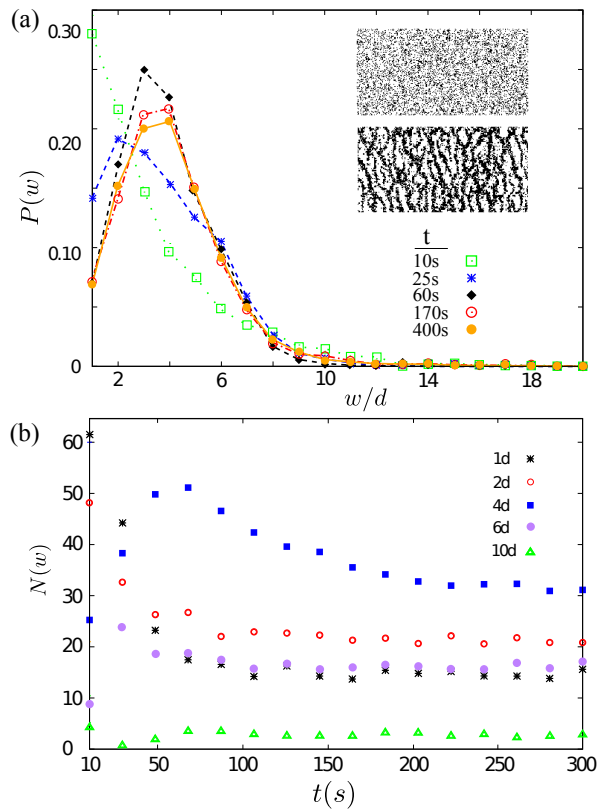}
\caption{(a) Evolution of the probability distribution 
$P(w)$ in MD simulations. Inset: Snapshots of particle 
positions, projected on the $xy$-plane, taken at $t{=}
\text{5s}$ (top) and $t{=}\text{300s}$ (bottom). (b) The 
average width-resolved number of stripes $N(w)$ along 
the $x$-axis as a function of time.}
\label{Fig8}
\end{figure}

Using this model, we carry out simulations that are 
able to reproduce the observed behavior in experiments (see 
Fig.~\ref{Fig8}). The cell is vibrated sinusoidally with 
frequency $f$ and amplitude $A$. The initial homogeneous 
state either remains well-mixed or quickly turns into a 
striped pattern, depending on the choice of the parameter 
values. Numerical simulations enable us to systematically 
determine the criteria under which the patterns form. For 
example, we find that the interparticle as well as wall-particle 
friction plays a crucial role in the formation of patterns. 
Indeed, no evidence for stripe formation was observed 
below $\mu{\sim}0.1$ in our simulations. Moreover, 
increasing the restitution coefficient of the collisions 
leads to nontrivial thickening of the stripes. The details 
of the extensive exploration of the parameter space for 
the pattern formation subdomain and their structural 
properties is beyond the scope of the present study and 
will be reported elsewhere. Here, we restrict the 
parameters to those of the experiments and focus on the 
structural characteristics of the resulting stripes. 
In contrast to the experiments, the initially formed 
distribution of the stripe width $P(w)$ at $A{=}5\bar r$ 
and $f{=}20 \,\text{Hz}$ is a monotonically decreasing 
function of $w$ which can be well fitted to an exponential 
decay [Fig.~\ref{Fig8}(a)]. The discrepancy can 
be attributed to the differences in the initial conditions 
and parameter values. The eventual form of $P(w)$, however, 
nearly follows a Poisson distribution with a peak at $w{\sim}4d\,
({=}8\bar r)$. Noticeably, the evolution of the width-resolved 
number of stripes $N(w)$ is also satisfactorily reproduced 
[Fig.~\ref{Fig8}(b)]. As a final remark, we note that 
varying the strength of the random noise (in the Langevin-type 
variant of the model) does not affect the probability 
distribution $P(w)$ significantly, while it clearly 
changes the average persistence length of the stripes. 
For very high noise strengths, the oscillatory drive is 
unable to generate order and the system remains in the 
mixed state. 

\section{Conclusion and outlook}
We have shown experimentally that the sinusoidal 
horizontal excitation of granular suspensions induces 
branched stripes aligned orthogonal to the direction 
of the periodic forcing. We characterized the structure 
by the probability distribution of the stripe thickness 
and verified that the structural changes evolve towards 
a dynamical saturation. It remains for further 
investigation to analyze the other structural features, 
such as spacefilling and branching properties; and also 
to clarify the role of key parameters on the formation 
and evolution of the patterns, aiming at understanding 
the underlying mechanisms and the possibility to design 
and control novel types of patterns. 

\begin{acknowledgments}
We thank H. Parishani of the University of Delaware for 
helpful discussions and comments and H. Pacheco for help 
with the experiments. R.M. and M.M. acknowledge the 
financial support by the IASBS Research Council under 
grant No.G2008IASBS136. 
\end{acknowledgments}

\end{document}